\documentclass[aps,prb,twocolumn,superscriptaddress,english,showpacs]{revtex4-1}

\usepackage{babel}
\usepackage{amsmath}
\usepackage{amssymb}
\usepackage{wasysym}
\usepackage{graphicx}
\usepackage[table]{xcolor}
\usepackage{multirow}
\usepackage[caption=false]{subfig}

\newcommand{\Graz}{Institute of Theoretical and Computational Physics, Graz University of Technology, NAWI Graz, 8010 Graz, Austria}
\newcommand{\Vienna}{Faculty of Physics, University of Vienna, 1090 Vienna, Austria}
\newcommand{\bbo}{BaBiO$_3$}
\newcommand{\monoclinic}{C12\={m}1}
\newcommand{\triclinic}{P\={1}}
\newcommand{\clustered}{C12\={m}1$^C$}
\newcommand{\distorted}{P1}
\newcommand{\icubic}{P3\={m}3}

\begin{document}

\title{Ab-initio prediction of the high-pressure phase diagram of \bbo}

\author{Andriy Smolyanyuk}       \affiliation{\Graz}
\author{Lilia Boeri}             \affiliation{\Graz}
\author{Cesare Franchini}        \affiliation{\Vienna}

\date{\today}

\begin{abstract}
{
\bbo\ is a well-known example of a 3D \textit{charge density wave} (CDW) compound, in which the CDW behavior is induced by charge disproportionation
at the Bi site.
At ambient pressure, this compound is a charge-ordered insulator, but little is known about its high-pressure behavior.
In this work, we study from first-principles the high-pressure phase diagram of \bbo\ using phonon modes analysis and evolutionary crystal structure prediction.
We show that charge disproportionation is very robust in this compound and persists up to 100~GPa. This causes the system to remain insulating up to the highest
pressure we studied.
} 
\end{abstract}



\maketitle

\section{Introduction}
\begin{figure}[b]
 \centering
   \begin{subfloat}[Pm\={3}m \label{fig:BBO_cubic}]{
   \includegraphics[width=0.33\linewidth]{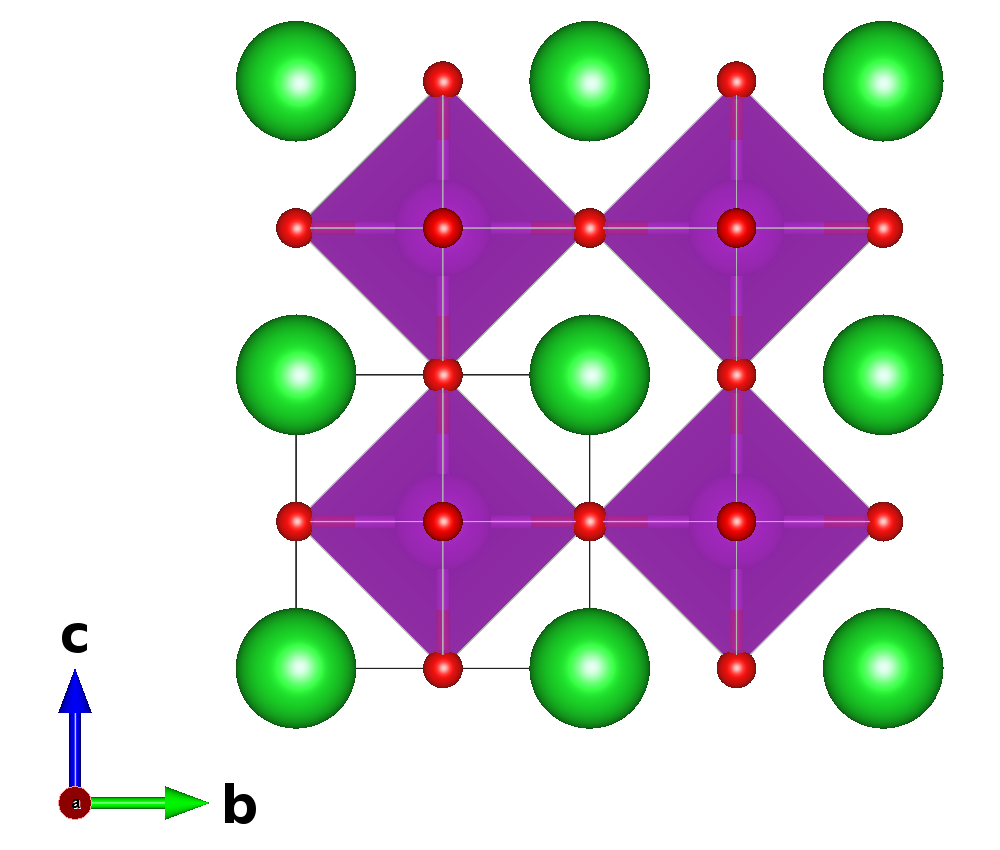}} 
   \end{subfloat}%
   \begin{subfloat}[C12/m1 \label{fig:BBO_mono}]{
   \includegraphics[width=0.33\linewidth]{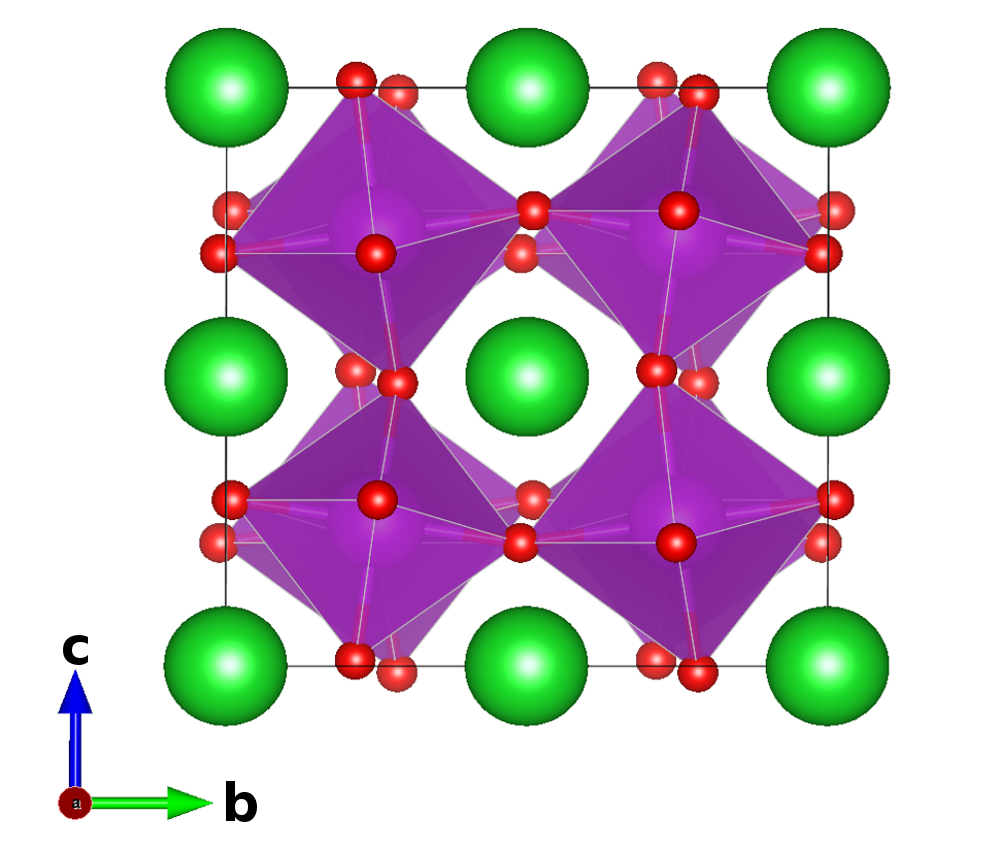}} 
   \end{subfloat}%
   \begin{subfloat}[P\={1} \label{fig:BBO_triclinic}]{   
   \includegraphics[width=0.33\linewidth]{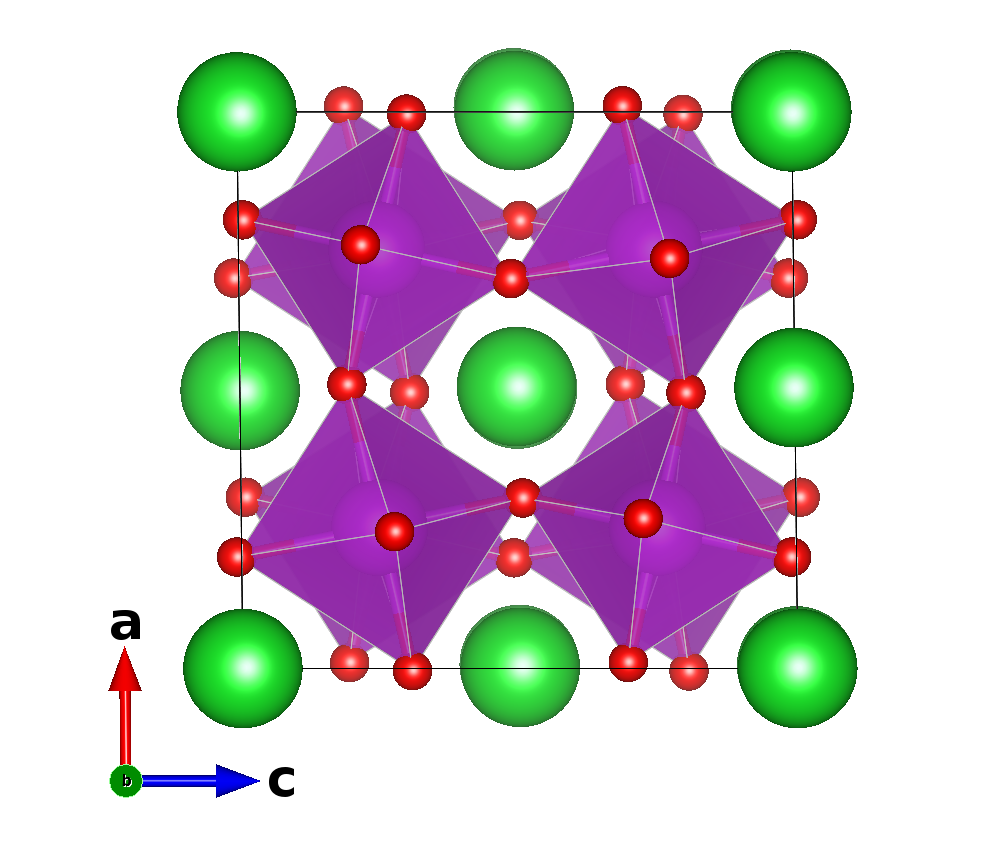}} 
   \end{subfloat}
   
   \begin{subfloat}[C12/m1$^C$ \label{fig:BBO_clustered}]{
   \includegraphics[width=0.58\linewidth ]{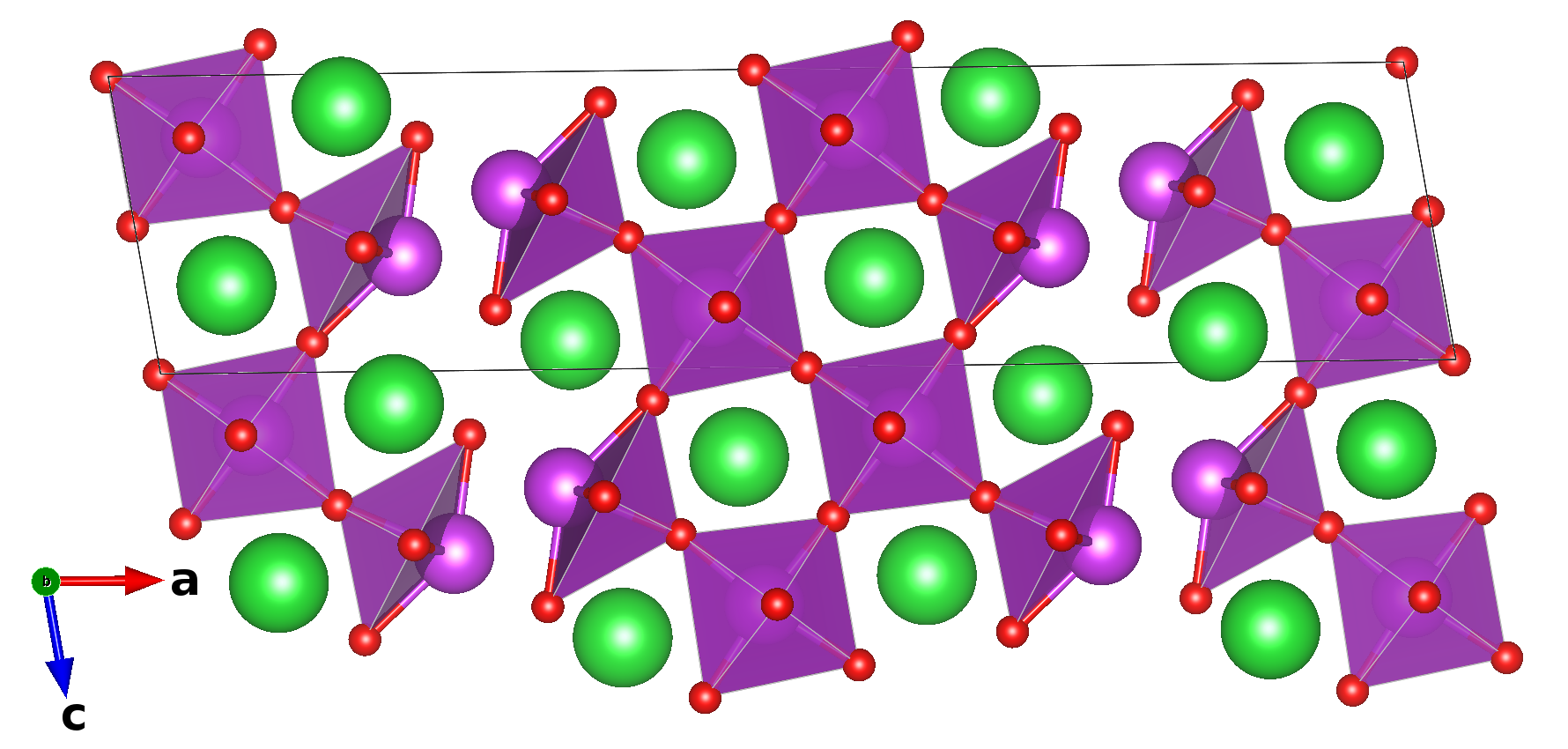}} 
   \end{subfloat}%
   \begin{subfloat}[Dist$_1$ \label{fig:BBO_dist1}]{   
   \includegraphics[width=0.4\linewidth]{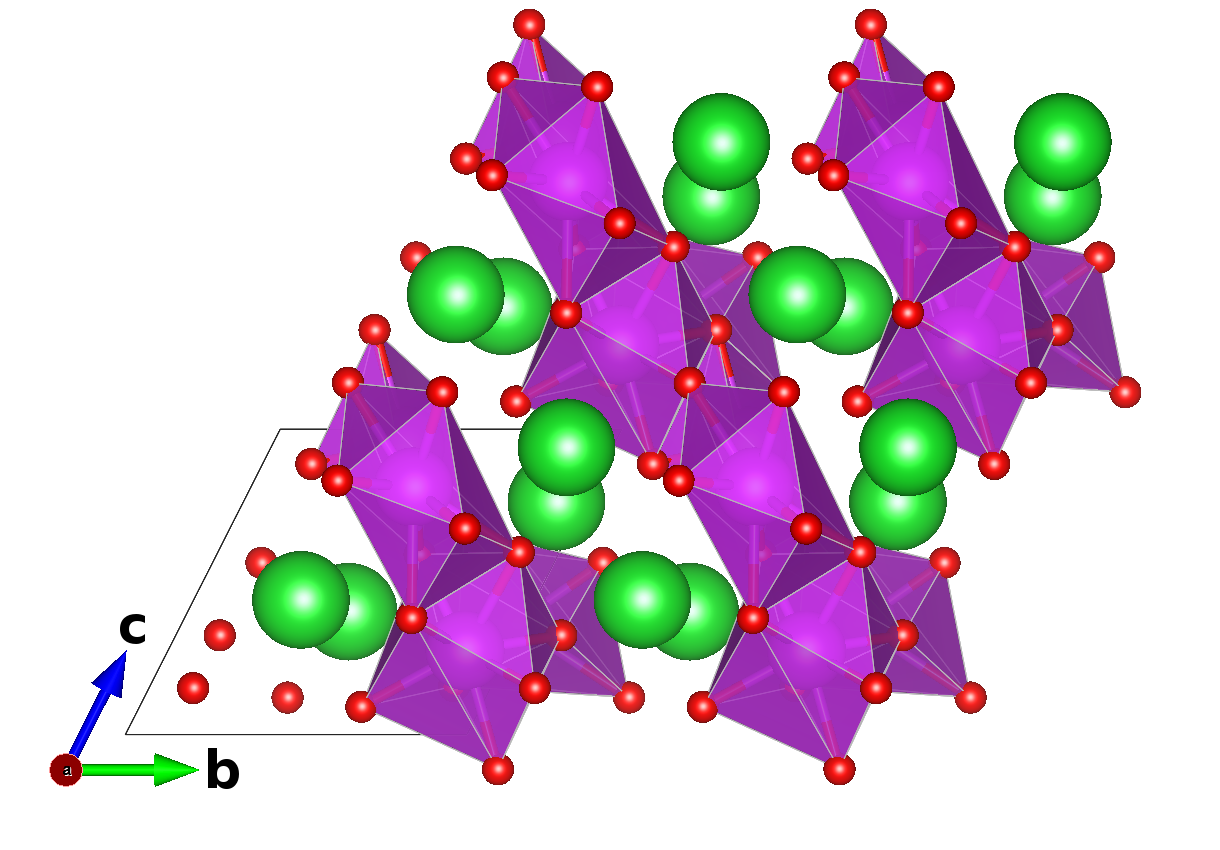}} 
   \end{subfloat}%
    \caption{({\em color online})
\bbo\ ground-state structures used in this work:
    (a)~ideal perovskite structure;
    (b)~experimental \bbo\ structure for ambient pressure
    and room temperature;
    (c)~triclinic structure stable from 20~GPa to 28~GPa pressure;
    (d)~\textit{``clustered''} monoclinic structure stable between 28~GPa and 87~GPa pressure;
    (e)~non-symmetric ``distorted$_1$'' structure stable from 87~GPa pressure.
    Structures (c), (d) and (e) where obtained using evolutionary crystal structure prediction (EP).
    Large green spheres are Ba atoms, large purple spheres are Bi atoms and small red spheres are O atoms.}  \label{fig:BBO_structures}
\end{figure}

Transition metal oxide perovskites are currently a subject of
intense study,
since they  exhibit many interesting properties which makes
them attractive for both fundamental and applied research.
Ferroelectricity \cite{ferroelectricity}, high-T$_c$ superconductivity, charge ordering, colossal magnetoresistance \cite{magnetoresistance}
are some of the most intriguing examples.
One of the most distinctive materials of this family is \bbo\cite{COX1976, bbo_review},
which is a charge-ordered insulator and becomes superconducting upon doping\cite{Uchida1987,Cava1988}.
Recent studies have shown that also at reduced dimensionality \bbo\ exhibits several remarkable properties:
experimental investigation of suppression of structural distortion in thin films \cite{suppresion_of_structural_distortion} and suppression of the \textit{charge density wave} (CDW) in the \bbo\ films
\cite{CDW_suppression}, as well as theoretical predictions of
topological phases\cite{topological_insulator} and  2D electron gas at the surface \cite{Surface_2DEG}  were recently reported in the literature.

In this compound $Bi$ behaves as mixed-valence element and can occur either in formal $Bi^{3+}$ or $Bi^{5+}$ 
oxidation state. The \bbo\ compound can thus be seen as a $ABB'O$ perovskite where $B$ is $Bi^{3+}$ 
and $B'$ is $Bi^{5+}$. Although the $Bi^{4+}$ oxidation state was never observed in any other compounds,
a hypothetical $BaBi^{4+}O_3$ should have an ideal perovskite cubic -- Pm\={3}m -- structure (Fig.~\ref{fig:BBO_structures}\subref{fig:BBO_cubic})
and should be a metal.
The mixed valence state of $Bi$ and the periodical arrangement of atoms leads to formation
of a CDW state.
This has two consequences: the different attraction strength of oxygen to the
two types of Bi leads to a chess-pattern breathing distortion of the Bi-O octahedra; in turn,
this CDW breathing distortion splits the antibonding Bi(s)-O(p) antibonding state and establishes an insulating
ground state\cite{Franchini2009, Sawatzky}.

Upon hole doping \bbo\ undergoes an insulator to metal transition and becomes a superconductor; superconducting T$_c$'s up to 13 K were reported for
 BaPb$_{1-x}$Bi$_{x}$O$_3$\cite{Uchida1987}, and up to 34~K for Ba$_{1-x}$K$_{x}$BiO$_3$\cite{Cava1988}.
The suppression of the CDW state upon doping, and the 
of superconductivity, have been intensely debated in literature.~\cite{Zeyher1990,Blaha1994,Liechtenstein1991,Kunc1991, Meregalli1998,Franchini2009,Franchini2010,Korotin2012,Bazhirov2013,Yin2013,Sawatzky}. Only recently, with the adoption
of post-DFT techniques, it was possible to obtain a consistent description of the whole phase diagram.

In many compounds, high pressure is a viable alternative to doping to 
tune the material properties. In particular, this has
recently been exploited in several 2D transition metal
dichalchogenides to study the interplay of CDW and superconductivity.
For instance, it was demonstrated that in pristine 1T-TiSe$_2$ \cite{Pressure_induced_SC_TiSe} and 2H-NbSe$_2$\cite{quantum_melting}
pressure induces a quantum melting of the CDW.
It is extremely intriguing to investigate whether a similar
behavior could occur in \bbo, i.e. whether pressure could be used to
suppress the three-dimensional CDW insulating state and
promote a metallic, superconducting state. 
Based on  simple theoretical arguments, 
all insulators 
should become metallic at high enough pressure (or density), 
due to increasing hybridization and higher bandwidth.
It is reasonable therefore to expect that, under pressure, \bbo\
 should undergo a structural transition to
the ideal perovskite structure, becoming metallic
by band overlap. 
However, this picture may be too naive.

In fact, in the last two decades the study of matter at extreme conditions
has shown that even 
simple elements exhibit a physical and chemical behavior, which cannot
be explained in these simple terms.
For example, simple free-electron metals such as lithium and sodium become
insulating under pressure;~\cite{lithium_hp,Ma2009} hydrides, which form insulating molecular crystals at
ambient pressure, can become metallic, and exhibit superconductivity with 
$T_c$'s as high
as 200~K;~\cite{Duan2014,Drozdov2015a,Drozdov2015b,Heil2015,Flores2016a,
Bernstein2015,Errea2015,Flores2016b,Shamp2016,Fu2016,Kokail2016}
even higher $T_c$'s have been predicted for elemental hydrogen, 
above the Wigner-Huntington transition.~\cite{Wigner1935,Cudazzo2008,Borinaga2016,Dias2017} Explaining the high-pressure behavior of solids thus requires first
an accurate understanding of the structural modifications induced by pressure.

The aim of this work is
to investigate the possibility of suppressing the CDW state and induce an insulator to metal transition in \bbo\ under pressure, performing a theoretical study of its crystal and electronic structure. 
Since the available experimental information is extremely limited,
we are using two structural prediction techniques to determine the possible high-pressure phases. The first is based on
group theoretical analysis of irreducible representations of unstable phonon modes using the \texttt{ISOTROPY} software suite \cite{Stokes1991} 
and the second is based on using evolutionary algorithms approaches as implemented in the \texttt{USPEX} package \cite{uspex}.
Our calculations show that charge ordering remains up to 100~GPa and the system remains insulating in this pressure range.

To investigate the electronic properties  we are using a hybrid Hartree-Fock/Density Functional Theory (DFT) calculation with the HSE (Heyd-Scuseria-Ernzerhof) exchange-correlation functional\cite{HSE06}.
In fact, the role of electronic correlations on the accurate description of the CDW was discussed in many papers and it was shown that
it is essential to use post-DFT approaches to obtain a proper  description of 
the electronic structure of this system \cite{Korotin2012,Yin2013,Franchini2009,Franchini2010}.

This paper is organized as follows: in Section \ref{sec:Results} we describe the results of our \textit{\textit{ab-initio}} calculations starting with our prediction for
the high-pressure phase diagram (Subsection~\ref{subsec:PredictedPhaseDiagram}), followed by the description of the phonon mode analysis
in Subsection~\ref{subsec:PhononModesAnalysis}.
We present the results obtained with \textit{ab-initio} evolutionary prediction method in Subsection~\ref{subsec:Ab-initioEveolutionaryPrediction} and 
electronic properties of the predicted ground-state structures are analyzed in Subsection~\ref{subsec:ElectronicProperties}. The main conclusions of this
work are summarized in Subsection~\ref{sec:Conclusions}.
Computational details are described at the end of the paper in Section~\ref{sec:ComputationalDetails}.

\begin{figure}[t]
 \includegraphics[width=\linewidth]{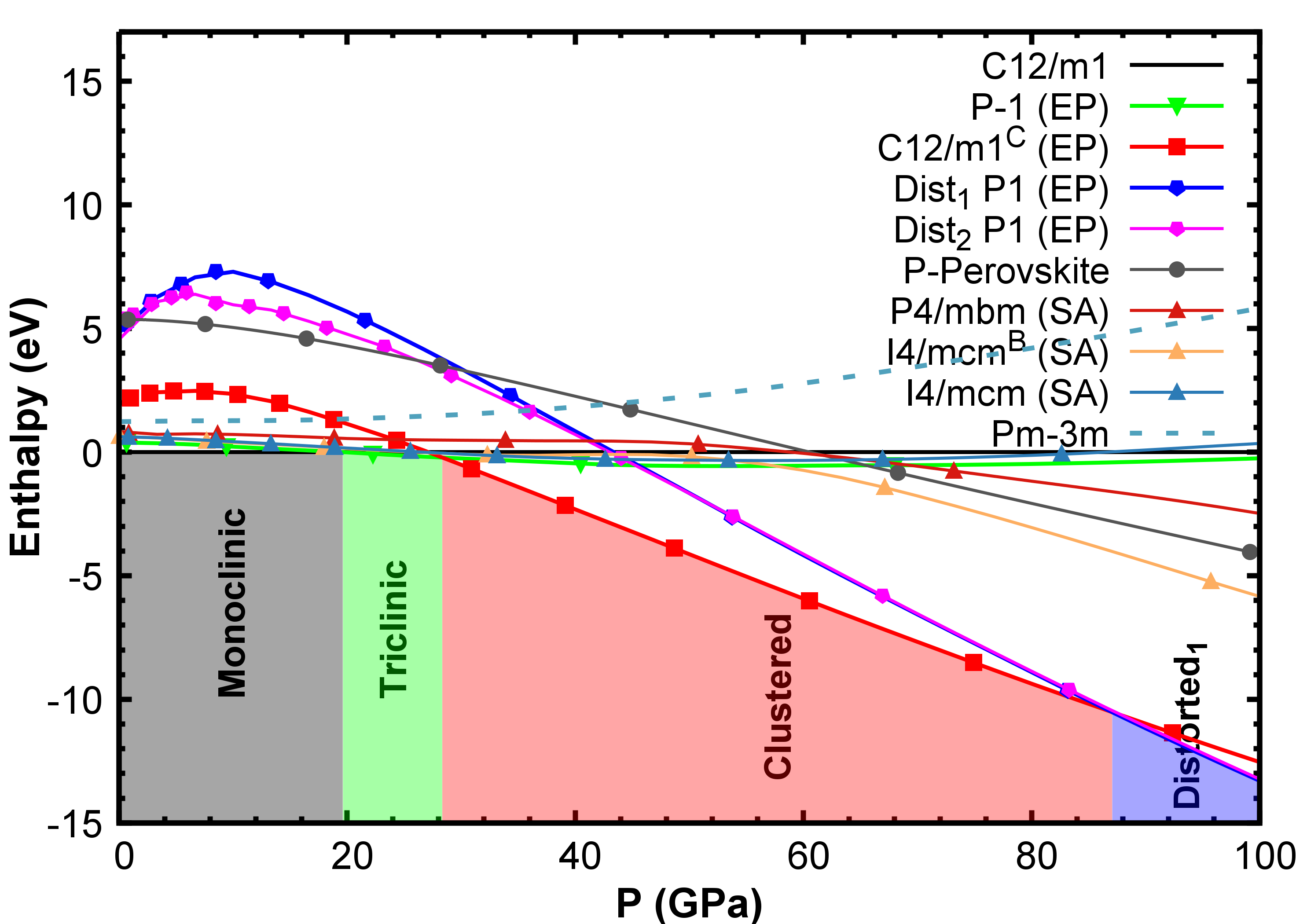}
 \caption{({\em color online})
Predicted high-pressure phase diagram of \bbo. Up to 100 GPa, we predict three structural phase transitions: monoclinic (\monoclinic) to triclinic (\triclinic) at about 20~GPa,
 triclinic to \textit{``clustered''} (\clustered) at about 28~GPa and \textit{``clustered''} to ``distorted$_1$'' (\distorted) at about 87~GPa. EP (Evolutionary Prediction) 
 means that structures were obtained
 using the evolutionary algorithms  structural prediction method  and SA (Symmetry Analysis) -- using the group-theoretical approach.}
 \label{fig:phase_diagram}
\end{figure}

\section{Results}
\label{sec:Results}
\bbo\ was synthesized for the first time  in 1963 \cite{BBO_prepared} and was
identified as a perovskite but it was hard to precisely determine the right structure: rhombohedral, orthorhombic, monoclinic or triclinic
symmetries were reported by various groups\cite{COX1976, triclinic, tetragonal}.
It was finally established that at room temperature and ambient pressure the structure is monoclinic (C12\={m}1)  -- Fig.~\ref{fig:BBO_structures}\subref{fig:BBO_mono}.
This structure can be described as a perovskite with two type of
distortions: \textit{breathing} and \textit{tilting} distortions of Bi-O octahedra.
In the breathing distortion the octahedra alternate in size, while upon tilting the octahedra are rotated.
The role of the breathing and tilting distortions has been widely discussed in literature. In particular, DFT calculations reported conflicting
results\cite{Zeyher1990,Liechtenstein1991,Blaha1994,Liechtenstein1991,Kunc1991}.
Only recently the use of post-DFT methods has permitted to reproduce the experimental findings correctly\cite{Franchini2009,Franchini2010,Korotin2012}.

\subsection{Predicted Phase Diagram}
\label{subsec:PredictedPhaseDiagram}

Fig.~\ref{fig:phase_diagram} shows our predicted high-pressure phase diagram 
\footnote{To model the effect of pressure on the system we have used volume associated to a given pressure  by doing fixed volume
relaxation using \texttt{VASP} and allowing atomic positions and unit cell shape to vary for several pressures.}.
The system undergoes three structural phase transitions  and remains insulating up to 100~GPa.
Fig.~\ref{fig:BBO_structures} shows the structures which are most relevant for the discussion.

As a reference structure we chose the monoclinic (\monoclinic) \bbo\ structure, which was determined by the experiment at the ambient pressure.
The structures with $EP$ after the title are the best structures obtained using evolutionary crystal structure prediction method.
They become energetically favorable in the high pressure region.
The structures with $SA$ are  obtained using the group-theoretical approach for crystal structure prediction. At high pressure these structures are energetically more favorable than 
our reference  (\monoclinic) structure, but are way behind  the structures obtained using the evolutionary method.
The {\em P-Perovskite} is a post-perovskite structure that we added into consideration by chemical intuition. This  structure was obtained by Oganov et al. for MgSiO$_3$ 
at extreme pressure and has 20 atoms in the unit cell \cite{PostPerovskite}. 

Following the lowest enthalpy path in our calculated phase diagram, we find that a
first transition from the monoclinic (\monoclinic, see Fig.~\ref{fig:BBO_structures}\subref{fig:BBO_mono}, 10 atoms in the unit cell)
to triclinic (\triclinic, see Fig.~\ref{fig:BBO_structures}\subref{fig:BBO_triclinic}, 40 atoms in the unit cell) structure occurs at about 20~GPa.
The \triclinic\ structure is a distortion of the \monoclinic\ structure with one additional tilting axis.
A transition from triclinic (\triclinic) to \textit{clustered} monoclinic (\clustered, Fig.~\ref{fig:BBO_structures}\subref{fig:BBO_clustered},
40 atoms in the unit cell) occurs
at about 28~GPa. We call this structure \textit{clustered} because it has ``domains'' created by shearing parts of the original structure.
The last transition is at about 87~GPa, where the \textit{clustered} monoclinic (\clustered) is transformed to highly distorted structure with no symmetry (\distorted, see
Fig.~\ref{fig:BBO_structures}\subref{fig:BBO_dist1}, 20 atoms in the unit cell). We will refer in text to it as \textit{Dist$_1$}. 

The experimental information about \bbo\ at high pressure is extremely scarce and limited to $P~\leq~20$~GPa\cite{sugiura_compression_1984, sugiura_high_1986}.
Unfortunately, the experimental resolution  was not enough to have a full refinement of crystal structure.
An anomaly at about 4~GPa was reported, at which the bulk modulus decreases
and this was attributed to the change of the tilt system (due to additional type of octahedra tilting).
Moreover it was reported that the system remains insulating up to at least 10~GPa.
Our results are consistent with the measurements that \bbo\ is an insulator in this pressure range. However, we cannot reproduce the anomaly at 4~GPa.

Before discussing in detail the crystal and electronic structure across the phase diagram, we will illustrate the details of our structural search methods.

\subsection{Phonon Mode Analysis}
\label{subsec:PhononModesAnalysis}

\begin{figure}[t]
 \centering
 \includegraphics[width=0.7\linewidth]{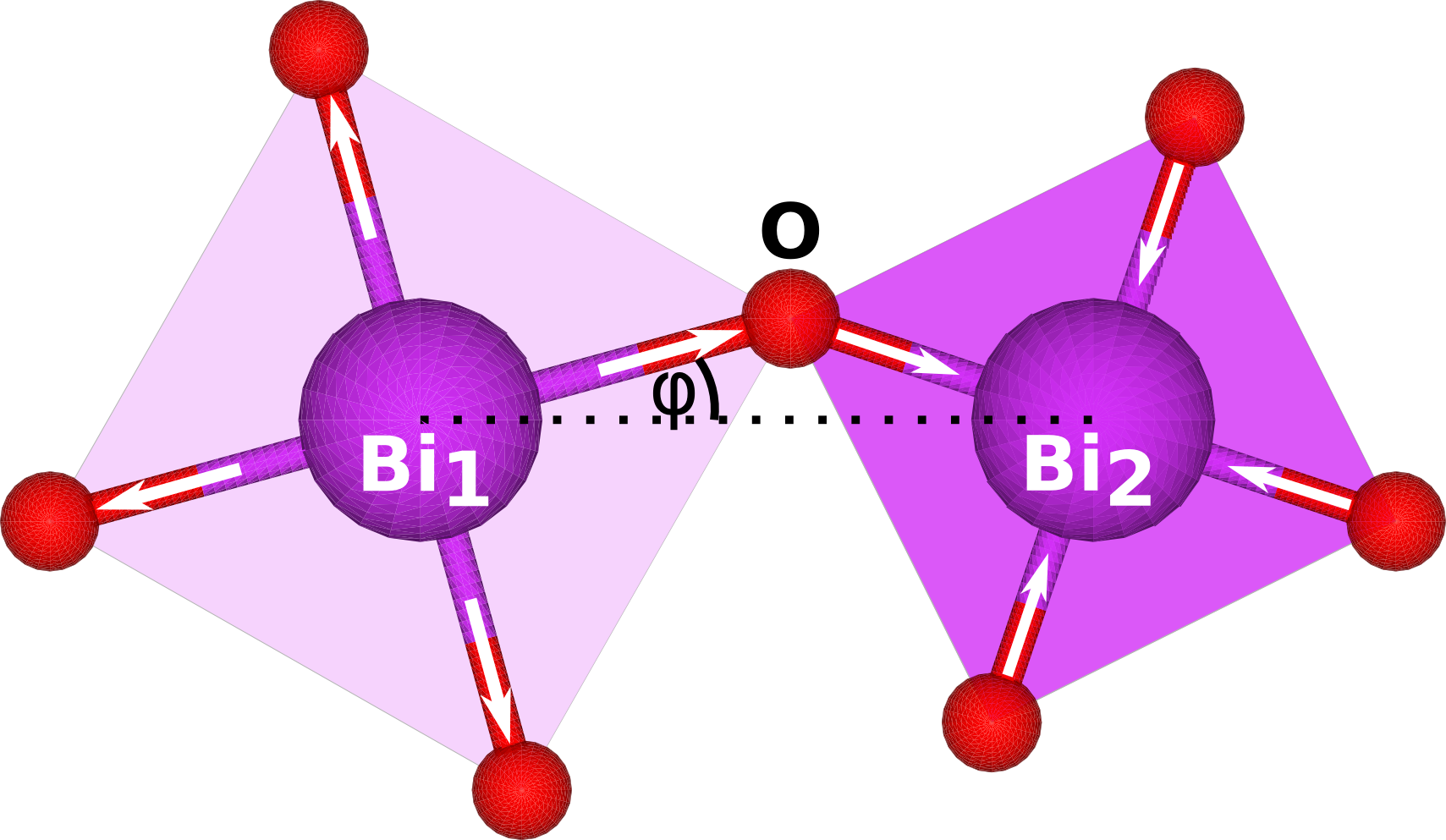}
 \caption{Breathing and tilting distortions applied to an ideal cubic perovskite structure.
 Breathing is indicated by squares of different size where the white arrows show the
 direction in which O atoms are displaced. Tilting is shown as a rotation of a square by angle $\varphi$ around axis normal to plane of paper.}
 \label{fig:scheme}
\end{figure}

%
%
%

\definecolor{tcG4m}{rgb}{1.00,0.00,0.00}
\definecolor{tcG5m}{rgb}{1.00,0.00,0.00}

\definecolor{tcM3p}{rgb}{0.00,0.63,0.00}
\definecolor{tcM3m}{rgb}{0.00,0.63,0.00}
\definecolor{tcM5m}{rgb}{0.00,0.63,0.00}

\definecolor{tcR1p}{rgb}{0.31,0.38,0.82}
\definecolor{tcR4p}{rgb}{0.31,0.38,0.82}

\definecolor{tcX5m}{rgb}{1.0,0.89,0.0}

%
%
%

{%
\begin{table}[b]
\centering
\caption{({\em Color online:}) Irreducible representations associated with unstable phonon eigenvectors of \icubic\ structure at 0~GPa and 100~GPa pressure.
``\#'' is a mode number.}
\label{table:irreps}

\bgroup
\def\arraystretch{1.5}

\begin{tabular}{|c|c|c|c|c|c|c|}
\hline
\multicolumn{3}{|c|}{\textbf{P=0 GPa}} & \multicolumn{4}{c|}{\textbf{P=100 GPa}} \\ \hline
\textbf{\#}  & \textbf{M} &  \textbf{R}  & $\mathbf{\Gamma}$ & \textbf{M} &  \textbf{R} & \textbf{X} \\ \hline
1&	\cellcolor{tcM3p}\textbf{M3+}	&	\cellcolor{tcR1p}\textbf{R1+}	&
\cellcolor{tcG4m}&	\cellcolor{tcM3p}\textbf{M3+}&		\cellcolor{tcR4p}&		\cellcolor{tcX5m} \\ 
\cline{1-3} \cline{5-5}
2&			&	\cellcolor{tcR4p}		&
\cellcolor{tcG4m}&	\cellcolor{tcM5m}	&	\cellcolor{tcR4p}&	\cellcolor{tcX5m}\multirow{-2}{*}{\textbf{X5-}} \\
\cline{1-1}\cline{7-7}
3&					&	\cellcolor{tcR4p}		&
\cellcolor{tcG4m}\multirow{-3}{*}{\textbf{$\Gamma$4-}}&		\cellcolor{tcM5m}\multirow{-2}{*}{\textbf{M5-}}&	\cellcolor{tcR4p}\multirow{-3}{*}{\textbf{R4+}}&\\
\cline{1-1}\cline{4-6}
4&					&	\cellcolor{tcR4p}\multirow{-3}{*}{\textbf{R4+}}	&
\cellcolor{tcG5m}&	\cellcolor{tcM3m} M3-					&
	&								\\ 
\cline{1-1}\cline{3-3}\cline{5-5}
5&					&						&
\cellcolor{tcG5m}&						&
	\multirow{3}{*}{}&						 \\ 
\cline{1-1}
6&				&	&			
\cellcolor{tcG5m}\multirow{-3}{*}{\textbf{$\Gamma$5-}}&		&
	&								\\ \hline
\end{tabular}

\egroup
\end{table}
}%

The group-theoretical method has been successfully used to identify and predict structural distortions in a variety of perovskites \cite{Howard, interplay_2013}. 
A powerful extension is to combine this method with the results of a phonon mode analysis of the parent structure \cite{Structural_determination}.
For \bbo\ the starting point is the ideal cubic perovskite structure -- Fig.~\ref{fig:BBO_structures}\subref{fig:BBO_cubic}.
It has a \icubic\ (cubic) symmetry with 5 atoms in the unit cell;
it is a metal, since no charge disproportionation is possible in this case.
The six O nearest neighbors of Bi form a perfect octahedron. This octahedron can be distorted in several ways and
there are two kinds of distortions usually present in real perovskite systems: \textit{breathing} and \textit{tilting} (see Fig.~\ref{fig:scheme}). 

A \textit{breathing} distortion means that some O atoms move closer or further away from the Bi atom, which is located at the center of the octahedron.
Octahedra are coupled to each other and the breathing of one of them creates an opposite breathing on its neighbors.
There are different kinds of breathing distortions depending on which and how many pairs of O are distorted.
In the monoclinic structure of \bbo\ all O atoms are involved in the breathing distortion and the overall picture is a 3D chess pattern of octahedra with alternating sizes.
A \textit{tilting} distortion is the rotation of an octahedron around a tilting axis. Neighboring octahedra in the plane normal to the tilting axis are coupled and rotate in the opposite direction.
Octahedra in adjacent planes may rotate in the same (\textit{in-phase} rotation) or in the opposite direction (\textit{antiphase} rotation).
The latter is present in  monoclinic \bbo.

The ideal cubic perovskite is dynamically unstable at ambient pressure and room temperature, and the analysis of unstable phonon modes gives an insight of 
which distortions correspond to a lowering of the energy of the system and thus allow to predict more favorable structures.
This can be achieved by computing the phonon dispersion relations and the corresponding eigenvectors. Eigenvectors with imaginary eigenvalues
represent the direction of atomic displacements that lead to a decrease in energy.
The stable structure can then be obtained following the most favorable distortion, which may be a linear combination of unstable phonon eigenvectors.
The space of possible combinations can be reduced using the group-theoretical method and the unstable phonon modes mapping on the corresponding irreducible
representations.

We calculated the phonon dispersion relations and the corresponding phonon eigenvectors for \bbo\ in the ideal cubic perovskite structure
\footnote{Results are available in the supplemental material}
using the \texttt{Phonopy} package \cite{phonopy} and classified unstable eigenvectors with the corresponding irreducible representations. 
We first tested our approach at ambient pressure, and the results we obtained are consistent with the experimental structure (see below).
After that, the investigation of stability of modes was repeated for several pressures in the range from 0 to 100~GPa (see Fig.~1 in supplemental material); 
the results for 0~GPa and 100~GPa pressure are summarized in Table~\ref{table:irreps}. 

At ambient pressure our result is consistent with the experimentally reported structure, namely monoclinic \bbo, since this structure can be obtained combining a  breathing  distortion 
with irreducible representation R1+ and the R4+ irreducible representation tilting  ($a^0b^-b^-$) distortion, which are both present in the results of our analysis.
The M3+ irreducible representation is an in-plane breathing distortion, which is not present in monoclinic \bbo\ because the R1+ breathing is more favorable.

\begin{table}
 \begin{center}
  \caption{\label{table:values}
 List of calculated by PBE and HSE values for monoclinic (\monoclinic) structure in comparison with experimental data.
$\delta$ is breathing distortion ($\delta=\frac{1}{2}(\overline{Bi_1O}-\overline{Bi_2O}$), where $\overline{Bi_1O}$ and $\overline{Bi_2O}$
are average Bi$_1$-O and Bi$_2$-O bond distances respectively), $\varphi$ is an average tilting angle and
$V$ -- volume per one formula unit of \bbo.}
\begin{ruledtabular}
  \begin{tabular}{cccc} 
 & $\delta$ (\r{A}) & $\varphi$ (deg) & V$_0$ (\r{A}$^3$)\\ \hline
PBE & 0.074 & 11.97 & 85.92\\ 
HSE & 0.096 & 10.95 & 82.67\\ 
PBE \cite{Franchini2010} & 0.07 & 12.1 & 85.76 \\
HSE \cite{Franchini2010} & 0.09 & 11.9 & 82.10 \\
Experiment \cite{COX1976} & 0.09 & 10.1  & 82.11 \\
\end{tabular}
\end{ruledtabular}

 \end{center}
\end{table}

In Table~\ref{table:values} we report the values of the breathing distortion ($\delta$), the tilting angle ($\varphi$) and equilibrium volume ($V_0$) for the ambient pressure
structure measured by experiment and calculated using both PBE (Perdew-Burke-Ernzerhof) exchange-correlation functional in the anzatz of the Generalized Gradient Approximation (GGA) and HSE functionals.
HSE gives a more accurate description of the structure, and reproduces correctly the electronic properties:
monoclinic \bbo\ is a charge ordered insulator
with a bandgap of about 0.8~eV, while in GGA it is a zero-gap semiconductor.
Although our calculations predict a transition from the monoclinic to the triclinic structure already at around 20~GPa, we followed the evolution of the monoclinic structure
to high pressure to see whether the monoclinic distortions are suppressed. We find that the distortions  are preserved under pressure and the tilting angle $\varphi$ is proportional to pressure.
Thus, the CDW is very robust in this compound and is not suppressed under pressure. Up to about 85~GPa the bandgap increases but afterwards its value decreases and the structure becomes a semi-metal
(the related information is provided in supplemental material in Fig.~2).

There are two regions were a substantial change in stability of the phonon modes takes place:
at about 30~GPa the R1+ mode become stable and at about 70~GPa the M3- and M5- modes become unstable (for the details see Fig.~1 in supplemental material).
We constructed our pool of structures using the unstable modes at 100~GPa, as this is the highest pressure we are interested in and unstable modes at lower pressures are a subset of these.
In particular, we have used the M3+, M5- and R4+  irreducible representations to construct our prediction for structures at high pressure (100~GPa) using a group-theoretical approach.
In total 17 structures were investigated with 10, 20 and 40 atoms in the unit cell (see Table~V in supplemental material for irreducible representations and order parameters used to obtain these structures).
The structures P4\={m}bm, I4\={m}bm and I4\={m}bm$^B$ (see the Fig.~\ref{fig:phase_diagram}) are the best obtained by this method followed by \texttt{VASP} relaxations.
P4\={m}bm is obtained from the ideal perovskite structure by applying in-phase tilting ($a^0a^0c^+$ in Glazer notation \cite{Glazer})
and I4\={m}bm by antiphase tilting ($a^0a^0c^-$). I4\={m}bm$^B$ is equivalent to I4\={m}bm with an additional in-plane breathing distortion. These three structures are metals, but
they are around 7 to 13.5~eV higher in enthalpy with respect to the ground state at 100~GPa, and thus very unlikely to be observed in experiments.

\begin{figure}[b]
 \centering
 \includegraphics[width=\linewidth]{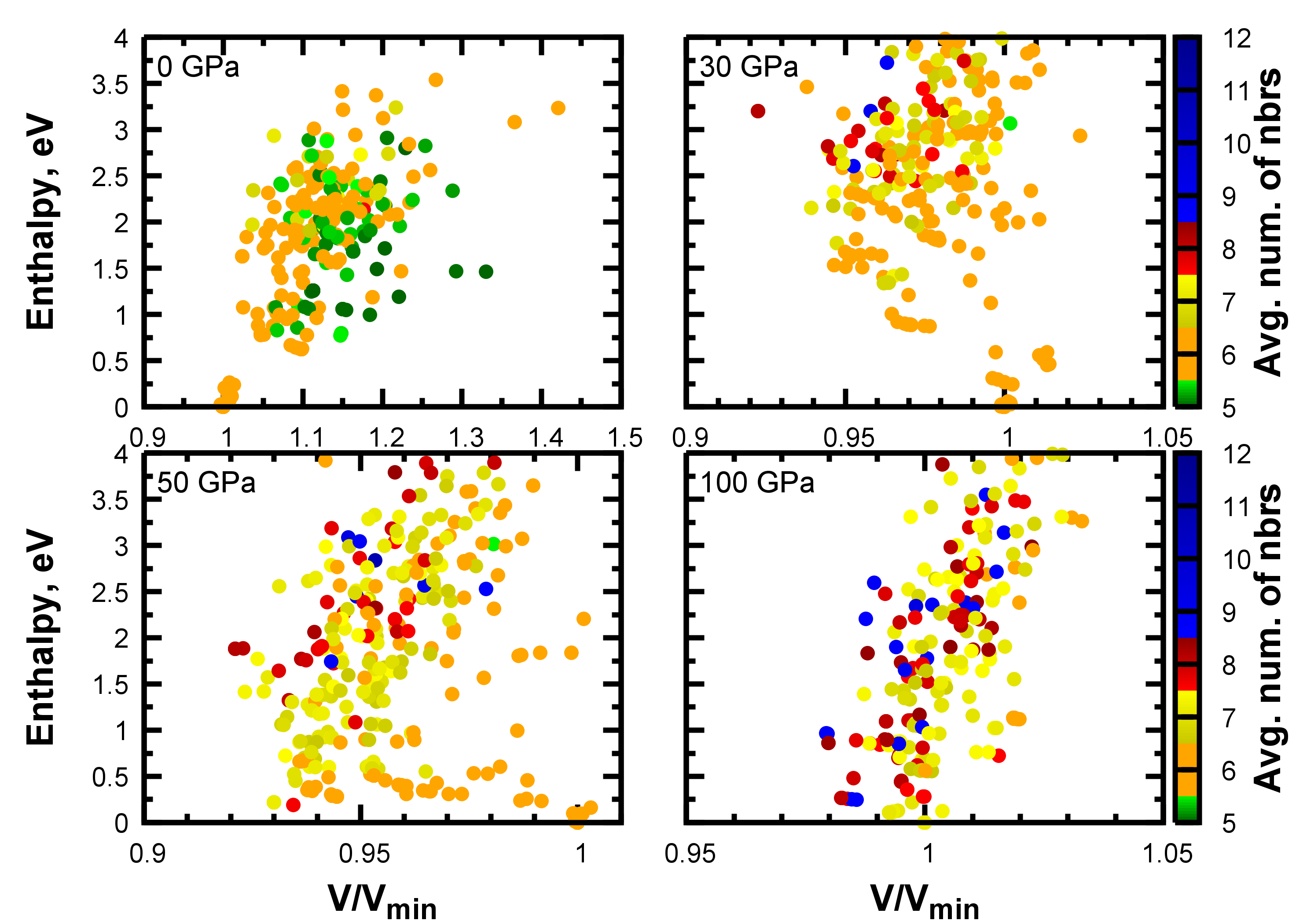}
 \caption{({\em Color online:})
Average number of O neighbors of Bi atoms for a pool of two formula units \bbo\  structures obtained using the evolutionary algorithms approach.
 Each dot is associated with a specific structure and its color denote the average number of neighbors. For the perovskite structure the average number of neighbors
 is 6 and the deviation from this value shows the amplitude of distortion presented in the system. Details on the neighbor analysis can be found in the supplemental material in subsection ``C. Neighbors analysis''.}
 \label{fig:Neighbours}
\end{figure}

\subsection{Ab-initio evolutionary prediction}
\label{subsec:Ab-initioEveolutionaryPrediction}

We performed ab-initio evolutionary algorithms calculations at fixed composition for structures with 10, 20 and 40 atoms in the unit cell
corresponding to 2, 4 and 8 formula units (f.u.) respectively.
The 10 atoms 0~GPa calculation was used to test the reliability of the method. The structure we obtained for this run is
consistent with the experiment. Calculations with 10, 20 and 40 atoms were made at 50 and 100~GPa. %
Up to 5-10 best structures from each run were chosen for more accurate relaxations to obtain the final ranking of the energies
and were relaxed for various pressures in the pressure range from 0 to 100~GPa using fixed volume relaxation 
allowing cell shape to change but with the symmetry fixed to obtain the equation of state.

The analysis of phonon modes for the monoclinic structure indicates a possible structural transition in the range of 25-30 GPa.
To check this region more carefully  we performed additional calculations at 25~GPa for structures with 20 and 40 atoms in the unit cell, and 
30~GPa  for structures with 10 atoms in the unit cell. As a result, we found a triclinic \triclinic\ structure that is stable in the 20-28~GPa pressure
region (Fig.~\ref{fig:BBO_triclinic}).
As already mentioned, the evolutionary algorithm approach was able to find the structures that are the energetically
most stable ones at high pressures (i.e. more stable than the best SA ones).
These are triclinic (\triclinic), \textit{``clustered''} (\clustered) and distorted (\distorted). 
There are both metals and insulators present in the pool of all structures obtained by these structural search calculations, but the most energetically favorable are always insulating
and strongly disordered.

The evolutionary algorithms calculations thus show that \bbo\ has a tendency to become more distorted when pressure is increased,
destabilizing the perovskite environment without suppressing the 
charge disproportionation.
In order to understand and visualize this tendency, we have performed an additional analysis on the pool of all 2 f.u. \bbo\ structures
predicted by the evolutionary algorithm approach.
The main idea of this analysis is to visualize the change in bonding environment induced by pressure, following the change in the average number of oxygen neighbors
for the Bi atoms. In an ideal perovskite this number is six; significant deviations from this value indicate strong distortions.
The result of this analysis  using the \texttt{Chemenv} module from \texttt{Pymatgen} package \cite{pymatgen, chemenv, Paper} is presented in Fig.~\ref{fig:Neighbours}
(for the details see subsection ``C. Neighbors analysis'' in supplemental material).
Each structure at the corresponding pressure is represented by a dot, with the color indicating the average number of neighbors of Bi and the coordinates describing the volume and enthalpy of the structure.
The volume is rescaled to the volume of the ground-state structure for a given pressure ($V_{min}$) and the enthalpy is given with respect to the enthalpy of the 
ground-state structure.
As expected, on average at ambient pressure there are six O atoms in Bi environment  and this number increases with
increasing pressure.
The plot thus shows that there is a general trend for \bbo\ to break the perovskite structure and become more distorted.

\subsection{Electronic properties}
\label{subsec:ElectronicProperties}

\begin{figure}[t]
 \centering
 \includegraphics[width=\linewidth]{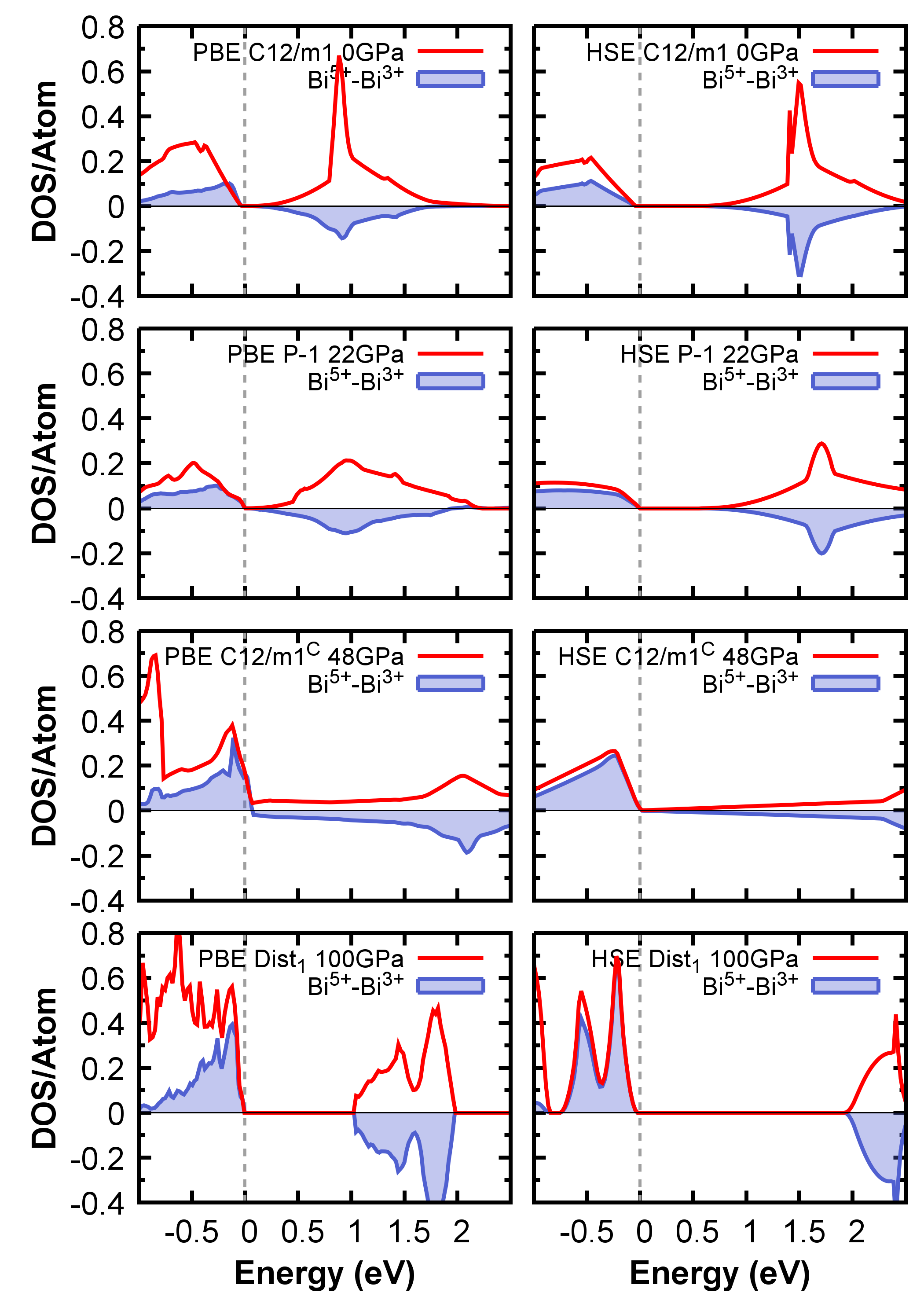}
 \caption{({\em Color online:}) DOS/atom and the difference of average PDOS for Bi$^{5+}$ and Bi$^{3+}$ atoms\cite{difference_note}
 for stable structures at the pressure of its stability. 
 Left panel shows results obtained with PBE functional for structural relations and DOS calculations.
 Right panel shows results for structures at the left panel but DOS calculations was performed using HSE functional using the structures from PBE calculations.}
 \label{fig:BBO_DOS}
\end{figure}

We analyzed the electronic properties of all four ground-state structures using both PBE and HSE. Our calculations show that 
all structures are insulating (see Fig.~\ref{fig:BBO_DOS}); the tendency is to become more insulating with
increasing pressure. Only the \clustered\ structure is metallic at the PBE level but on the HSE  level  opens a gap.
Here the use of HSE functional is important as PBE underestimates the band gap or predicts the structure to be
a metal while actually it is an insulator\cite{Franchini2010}. Analyzing the partial DOS for Bi atoms it may be seen that all the structures have two
inequivalent Bi atoms with different formal valence state, and the insulating behavior is associated with the CDW.

\begin{figure}[t]
 \centering
 \includegraphics[width=\linewidth]{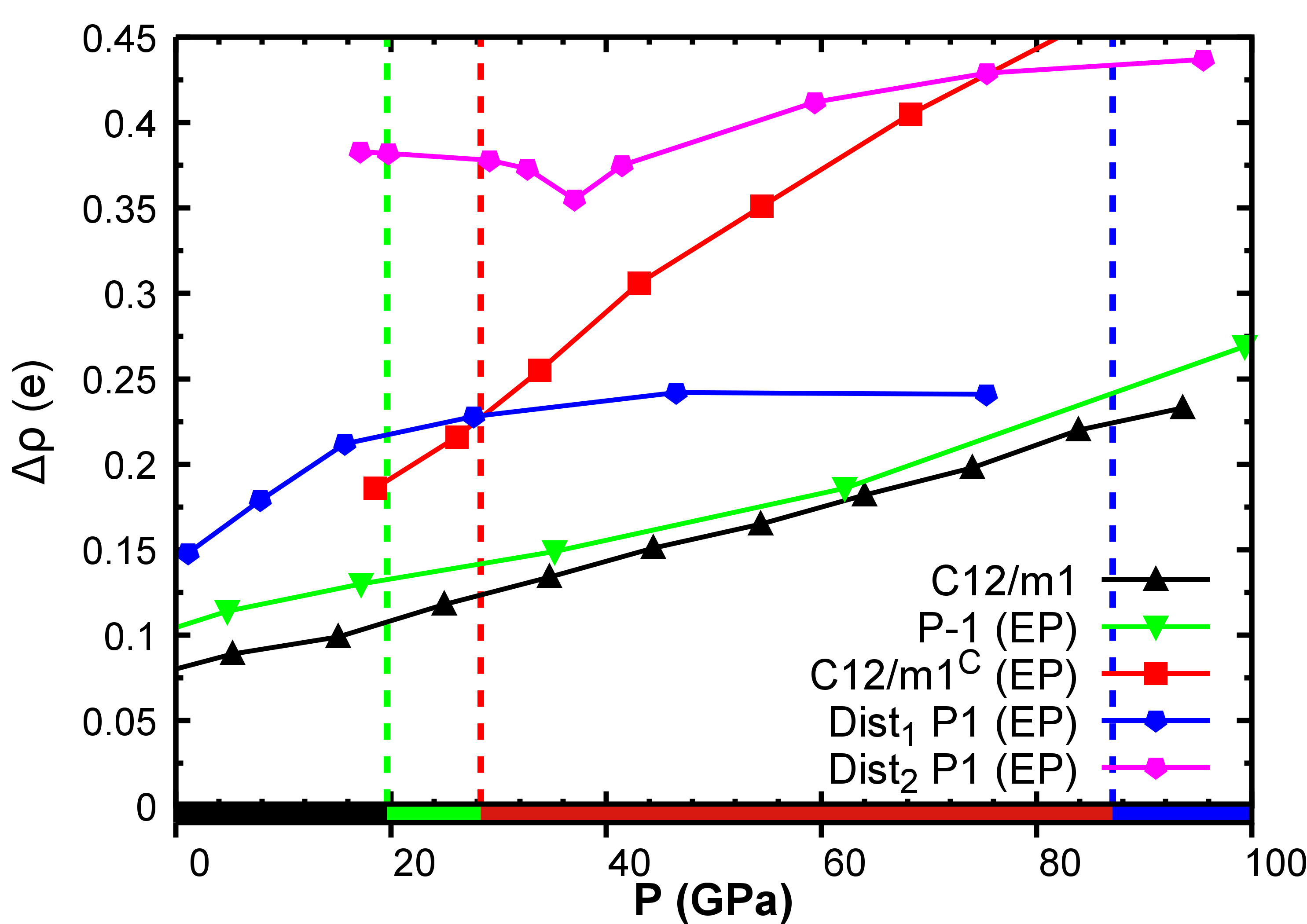}
 \caption{({\em Color online:})
Charge difference $\Delta \rho$ between
 two different Bi atoms for all stable structures in the phase diagram: the one with the maximal charge (represent Bi$^{3+}$ atom)
 and the one with the minimal (represent Bi$^{5+}$) -- HSE functional was used.
The color scale at the bottom indicates the stability range of the various structures.}
 \label{fig:charges}
\end{figure}

We determine the magnitude of the charge disproportionation by calculating the charge difference $\Delta \rho$  for the two types of
bismuth, Bi$^{3+}$ and Bi$^{5+}$, from the partial \textit{spd-charge} of the site projected ground-state wave function -- see Fig.~\ref{fig:charges}.
Note that, although we name these atoms Bi$^{3+}$ and Bi$^{5+}$, the actual difference in charge is much smaller than two\cite{Franchini2010}.

For the \monoclinic, \triclinic  \ and \clustered \  structures the charge disproportionation increases when the pressure is increased.
The change of charge disproportionation for Dist$_1$ and Dist$_2$ structures in the pressure range from 20~GPa to 100~GPa is small. Here Dist$_2$ is another distorted structure consisting of 8~f.u. obtained by the evolutionary algorithms approach.
Dist$_1$ and Dist$_2$ are competing phases at high pressure, but Dist$_1$  is always lower in energy.

The only possible way to suppress the CDW and have a metallic \bbo\ phase is to have a highly symmetric structure with no crucial distortions as electronic properties of \bbo\ are
highly coupled to the structural ones. 
Our calculations show that for \bbo\ all energetically favorable structures are distorted in a way they are forced to be insulators, and metallic (symmetric) structures
are strongly unfavorable in energy, and thus very unlikely to occur. 
 
\section{Conclusions}
\label{sec:Conclusions}
In this work we performed a  theoretical study of the \bbo\ phase diagram under pressure to analyze the possibility of suppressing the CDW distortion,
caused by charge disproportionation.
We used  two different structural search approaches to construct our \textit{ab-initio} predicted high-pressure phase diagram: evolutionary algorithms and phonon mode analysis.
The resulting phase diagram shows three structural phase transitions.
The group-theoretical structure prediction method alone is not able to find the best structures at high pressure as it is constrained only to structures with symmetry.
By using evolutionary algorithms we found that \bbo\ becomes more distorted with increasing pressure.
In fact, our calculations show that high pressure favor complicated structures with clustering and shearing distortions, which cannot be described
as symmetry distortions of the single perovskite structures.
The analysis of the electronic properties show that all ground-state structures remain insulating up to 100~GPa and the charge disproportionation is preserved at high pressure.
This hinders the transition towards a metallic regime.

\section{Computational details}
\label{sec:ComputationalDetails}
For total energies and structural optimization we used GGA and HSE DFT calculations, as implemented in the \texttt{VASP} package \cite{VASP1, VASP2, VASP3, VASP4}
using PAW pseudopotentials \cite{PAW1, PAW2}.
We used the hybrid HSE functional to improve the description of the electronic properties.
The energy cutoff value was set to 500~eV and $\Gamma$-centered Monkhorst-Pack grid \cite{MP1, MP2} with 4x4x4 k-points was used for the GGA \cite{GGA1, GGA2} and HSE 
functional for the structural relaxation.
To do the relaxation at a specific pressure we performed fixed-volume calculation.
For DOS calculations we used 8x8x8 grid for GGA functional and 2x2x2 for HSE which was sufficient to converge the value of the semiconducting gap (note that all structures at HSE 
level are semiconducting).
Phonon calculations were done with 20x20x20 k-points grid at PBE level.

We have used two approaches to predict possible structures at high pressure: evolutionary algorithms as implemented in
\texttt{USPEX} package \cite{uspex} and the group-theoretical analysis using \texttt{ISOTROPY} software suite \cite{Stokes1991}. Determination of dispersion relations were done with the \texttt{Phonopy} package\cite{phonopy}.
The setup for \texttt{USPEX} calculations was the following: 20 structures in population,  40 structures in the initial population, maximal number of generations is 25 and stopped the search when the best structure did not change for 8 generations.

The analysis of the Bi environment for the pool of structures obtain in the \texttt{USPEX} calculations was performed using  the \texttt{Chemenv} module from the \texttt{Pymatgen} 
package \cite{pymatgen, chemenv, Paper}.
The main goal of the \texttt{Chemenv} module is to determine the chemical environment of each atom in a structure finding the best polyhedron that can 
represent atomic positions using \textit{continuous symmetry measure} as a parameter to determine the optimal polyhedron (more details in the supplemental material).
The pool of structures consists of around 300 structures for each pressure with 10 atoms in the unit cell.

\begin{acknowledgments}
We thank Jiangang He for useful discussions and helpful instructions on symmetry analysis
and David Waroquiers and Geoffroy Hautiers for the development of the \texttt{chemenv} package and very helpful technical support on its use.
We acknowledge funding from the Austrian Science Fund FWF through SFB ViCoM, Project F41-P15 and 
computational resources from the VSC3 of the Vienna University of Technology.
\end{acknowledgments}

\bibliographystyle{apsrev4-1}
\bibliography{paper}

\begin{thebibliography}{70}
\bibitem [{dif()}]{difference_note}%
  \BibitemOpen
  \href@noop {} {}\bibinfo {note} {It is possible to precisely determine the
  difference of partial DOS only for monoclinic \monoclinic\ \bbo\ as it has
  only two Bi atoms which are inequivalent. For the other structures we have
  separated Bi atoms into two formal groups representing Bi$^{5+}$ and
  Bi$^{3+}$ case. After averaging partial DOS in each group the difference was
  taken.}\BibitemShut {Stop}%
\end{thebibliography}

\end{document}